\magnification=\magstep1
\overfullrule=0pt
\parskip=6pt
\baselineskip=15pt
\headline={\ifnum\pageno>1 \hss \number\pageno\ \hss \else\hfill \fi}
\pageno=1
\nopagenumbers
\hbadness=1000000
\vbadness=1000000

\input epsf
%----------------------------------------------------------------------------

\vskip 25mm
\vskip 25mm
\vskip 25mm

\centerline{\bf ON POINCARE POLINOMIALS OF HYPERBOLIC LIE ALGEBRAS}
\vskip 15mm

\centerline{\bf H. R. Karadayi}
\centerline{Dept. Physics, Fac. Science, Istanbul Tech. Univ.}
\centerline{80626, Maslak, Istanbul, Turkey }
\centerline{e-mail: karadayi@itu.edu.tr}
\centerline{and}
\centerline{Dept. Mathematics, Fac. Science, Istanbul Kultur University}
\centerline{34510, Sirinevler, Istanbul, Turkey }

\vskip 5mm

\centerline{\bf M. Gungormez}
\centerline{Dept. Physics, Fac. Science, Istanbul Tech. Univ.}
\centerline{80626, Maslak, Istanbul, Turkey }
\centerline{e-mail: gungorm@itu.edu.tr}

\vskip 5mm
\medskip

\centerline{\bf{Abstract}}

Poincare polinomials of hyperbolic Lie algebras, which are given by $HA_2$ and $HA_3$ 
in the Kac's notation, are calculated explicitly. 
The results show that there is a significant form for hyperbolic Poincare polinomials. Their explicit forms tend to be seen as the ratio of a properly chosen finite Poincare polinomial and a polinomial of finite degree. To this end, by choosing the Poincare polinomials of $D_4$ and $D_5$ Lie algebras, we show that these polinomials come out to be of order 11 and 19 respectively for $HA_2$ and $HA_3$.

\vskip 15mm
\vskip 15mm
\medskip

\hfill\eject

\vskip 3mm
\noindent {\bf{I.\ INTRODUCTION }} 
\vskip 3mm

To be the central idea of this work, a characteristic fact about Poincare polinomial 
$P(G_r)$ of a Kac-Moody algebra $G_r$ of rank r {\bf [1]} is that the term of order n 
gives the number of its Weyl group elements which are composed out of the products of n
Weyl reflections corresponding to simple roots of $G_r$ {\bf [2]}. Poincare polinomials
are already known {\bf [3]} in the following form

$$  P(G_r) \equiv \prod^r_{i=1}  { t^{d_i}-1 \over t-1}  \eqno(I.1) $$

\noindent for a finite Lie algebra $G_r$ where $d_i$'s are the degrees of its basic 
invariants and t is an indeterminate. For an affine Kac-Moody algebra originated from 
a generic finite Lie algebra $G_r$, Bott theorem states that its Poincare polinomial 
has the form
 
$$ P(G_r) \prod^r_{i=1} {1 \over 1-t^{d_i-1}} .  \eqno(I.2) $$
  
Very little is known however beyond Affine Kac-Moody algebras. At least for Hyperbolic Lie Algebras, our observation is that their Poincare polinomials come in the form

$$ { P(G_r) \over P_N(t) }   \eqno(I.3) $$

\noindent where $G_r$ is a properly chosen finite Lie Algebra and $P_N(t)$ is a polinomial 
of some degree N in indeterminate t. At least for Hyperbolic Lie algebras of type $HA_r$, 
our study concerning (I.3) tells us that

\noindent (1) $G_r$ must be chosen as the maximal finite Lie algebra which is contained in the
Dynkin diagram

\noindent (2) On the contrary to Affine cases, $P_N(t)$ is not fully factorizable in indeterminate t.

\hfill\eject

Let us consider $HA_2$ Dynkin diagram as in the following form

%\midinsert
\epsfxsize=4cm
\centerline{\epsfbox{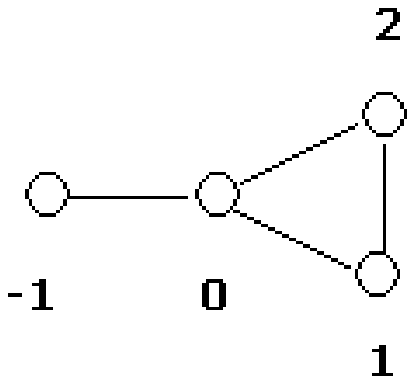}} \rightline{(I.4)}
%\endinsert
  
\noindent for which one has 

%\midinsert
\epsfxsize=3.5cm
\centerline{\epsfbox{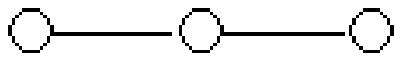}}
%\endinsert

\noindent by deleting the node 2,

%\midinsert
\epsfxsize=4cm
\centerline{\epsfbox{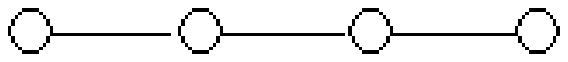}}
%\endinsert

\noindent by deleting the line between 0 and 2 and also

%\midinsert
\epsfxsize=3.5cm
\centerline{\epsfbox{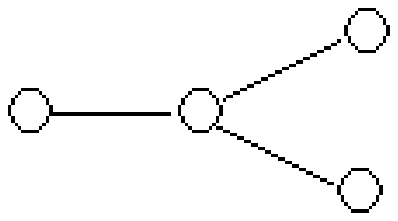}}
%\endinsert

\noindent by deleting the line between 1 and 2. This hence leaves us with three 
possibilities $A_3$, $A_4$ and $D_4$ to choose $P(G_r)$ in (I.3) and one succeeds
to find a $P_N(t)$ in all three cases respectively for N=5,9,11.

$HA_3$ diagram, on the other hand, has the following form

%\midinsert
\epsfxsize=4cm
\centerline{\epsfbox{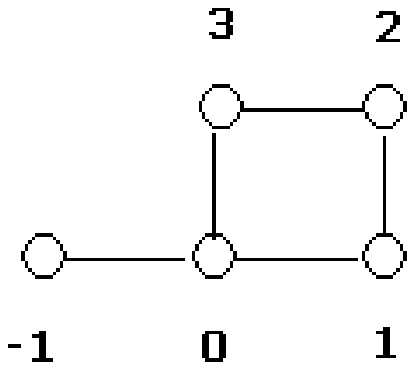}} \rightline{(I.5)}
%\endinsert

\noindent and in the same way as above, one obtains three possibilities $A_4$, $A_5$ 
and $D_5$ for which only $D_5$ leads us to a finite polinomial $P_{19}(t)$ of finite
order.

\vskip 3mm
\noindent {\bf{II.\ CALCULATION OF HYPERBOLIC POINCARE POLINOMIALS }}
\vskip 3mm

We follow the notation of Kac-Moody-Wakimoto {\bf [4]} for hyperbolic Lie algebras.
For $HA_3$ with the Dynkin diagram given in (I.5), simple roots $\alpha_\mu$ and
fundamental dominant weights $\lambda_\mu$ are given by equations
$$ \kappa(\lambda_\mu,\alpha_\nu) = \delta_{\mu,\nu} \ \ \ \ \ \  \mu,\nu = -1,0,1,2,3. $$
where $\kappa( ,)$ is the symmetric scalar product defined on $HA_3$ weight lattice by 
its Cartan matrix A with inverse

$$ A^{-1} = - \pmatrix{
0&& 1&& 1&& 1&& 1 \cr
1&& 2&& 2&& 2&& 2 \cr
1&& 2&& 5/4&& 3/2&& 7/4 \cr
1&& 2&& 3/2&& 1&& 3/2 \cr 
1&& 2&& 7/4&& 3/2&& 5/4 \cr} $$

\noindent so we have
$$ \lambda_{\mu} = \sum_{\nu=-1}^3 (A^{-1})_{\mu,\nu} \ \alpha_\nu $$

Let $G_r$ be a chosen Kac-Moody algebra, $ {\cal W}(G_r)$ be its Weyl group
and $\rho$ its Weyl vector. For any $\Sigma \in {\cal W}(G_r)$, let us now consider
$$ \Gamma \equiv \rho-\Sigma(\rho) \eqno(II.1)$$
which is by definition an element of the positive root lattice of $G_r$. In another 
work {\bf[5]}, we have shown that

\noindent (1) $\Gamma$ is unique in the sense that  $\Gamma \equiv \rho-\Sigma(\rho)$ 
is different from $\Gamma^\prime \equiv \rho-\Sigma^\prime(\rho)$ for any two different
$\Sigma$, $\Sigma^\prime \in {\cal W}(G_r)$,

\noindent (2) $\Gamma$ has a unique decomposition over root system of $G_r$, the set of 
roots with minimal square roots, with coefficients taking only the values 0 or 1.

\noindent The first one of these propositions could be easily understood due to definition 
of Weyl vector though the second one is somewhat non-trivial leading us to specify the
signatures of Weyl reflections without using their generic definitions which are quite hard 
to handle.

Within the scope of this work, the first one is sufficient to suggest our simple method to 
calculate the number of Weyl group elements which are expressed in terms of the same number
of simple Weyl reflections $\Sigma_\mu$ which are defined by 
$$ \Sigma_\mu(\Lambda) \equiv \Lambda - 2 \ 
{ \kappa(\Lambda,\alpha_\mu) \over \kappa(\alpha_\mu,\alpha_\mu) } \ \alpha_\mu  \ \ , \ \ 
\mu = -1,0,1,2,3  $$
for any element $\Lambda$ of weight lattice. Let us now consider products
$$  \Sigma_{\mu_1} \Sigma_{\mu_2} \dots \Sigma_{\mu_i}   \eqno(II.2) $$
which can not be reduced into products consisting less than i-number of simple Weyl reflections. Out of all these non-reducible elements as in (II.2), we define a class 
${\cal W}^i$ of ${\cal W}(G_r)$ and, by the aid of (II.1), $\Gamma^i$ can always be 
defined as being in correspondence with ${\cal W}^i$. We can formally state that any 
Weyl group is the formal sum of its classes ${\cal W}^i$. The number of classes is finite 
for finite Lie algebras and infinite for infinite Kac-Moody algebras. One should however 
note that the number of elements of ${\cal W}^i$ is always finite.

Looking back to $HA_3$, we note that elements  $ \gamma^i \in \Gamma^i$ are defined for any 
$\Sigma^i \in {\cal W}^i$ by $ \gamma^i \equiv \rho-\Sigma^i(\rho)$ where
$ \rho \equiv \lambda_{-1}+\lambda_0+\lambda_1+\lambda_2+\lambda_3 $. As is emphasized 
above, elements of $\Gamma^i$ are different from each other. We know that,
$$ \Gamma^0 \equiv \{ 0 \} \ \ \ , \eqno(II.3) $$
$$ \Gamma^1 \equiv \{ \alpha_{-1}, \alpha_0, \alpha_1, \alpha_2, \alpha_3 \}   
\eqno(II.4) $$
and the ones which are chosen to be different from the set
$$ \rho-\Sigma_\mu(\rho-\gamma^1) \ \ \ , \ \ \  \gamma^1 \in \Gamma^1  \eqno(II.5) $$
are the elements of $\Gamma^2$. For the present case one thus has
$$ \eqalign{ \Gamma^2 = \{
&2 \ \alpha_{-1} + \alpha_0 \ , \
\alpha_{-1} + 2 \ \alpha_0 \ , \
\alpha_{-1} + \alpha_1 \ , \
\alpha_{-1} + \alpha_2 \ , \ 
\alpha_{-1} + \alpha_3 \ , \  \cr
&2 \ \alpha_0 + \alpha_1 \ , \  
\alpha_0 + 2 \ \alpha_1 \ , \  
\alpha_0 + \alpha_2 \ , \     
2 \ \alpha_0 + \alpha_3 \ , \
\alpha_0 + 2 \ \alpha_3 \ , \  \cr
&2 \ \alpha_1 + \alpha_2 \ , \
\alpha_1 + 2 \ \alpha_2 \ , \
\alpha_1 + \alpha_3 \ ,  \ 
2 \ \alpha_2 + \alpha_3 \ , \
\alpha_2 + 2 \ \alpha_3 \} . }   $$

Elements of $\Gamma^i$ will be similarly chosen to be the different ones from the set
$$ \rho-\Sigma_\mu(\rho-\gamma^{i-1}) \ \ \ , \ \ \  \gamma^{i-1} 
\in \Gamma^{i-1}  \eqno(II.5) $$
and this leads us to the polinomial $ \sum_{i=0}^\infty \ dim {\cal W}^i \ t^i $
which is nothing but the Poincare polinomial of $HA_3$ algebra where $dim {\cal W}^i$ 
is the number of elements of $ {\cal W}^i$. By explicit calculation up to 27th order, 
we obtained the following result
$$ \eqalign{ P(HA_3) \equiv 
&1 + 5 \ t^1 + 15 \ t^2 + 36 \ t^3 + 75 \ t^4 + 142 \ t^5 + 252 \ t^6 + 428 \ t^7 +
704 \ t^8 + 1132 \ t^9 + \cr
&1791 \ t^{10} + 2800 \ t^{11} + 4339 \ t^{12} + 6680 \ t^{13} + 10234 \ t^{14} + 
15621 \ t^{15}  +  23778 \ t^{16} +   \cr
&36119 \ t^{17} + 54779 \ t^{18} + 82981 \ t^{19} + 125590 \ t^{20} + 
189949 \ t^{21} + 287142 \ t^{22} + \cr
&433899 \ t^{23} + 655471 \ t^{24} + 989971 \ t^{25} + 
1494923 \ t^{26} + 2257149 \ t^{27} + \dots } \eqno(II.6) $$ 
One sees that (II.6) is enough to conclude that 
$$ P(HA_3) \equiv { P(D_5) \over P_{19}(t) } \eqno(II.7) $$
where
$$ P_{19}(t) \equiv (1 + t^4) \ (1 - t^2 - t^3 - 2 \ t^4 - t^5 + t^7 + 3 \ t^8 + 2 \ t^9 + 
2 \ t^{10} + t^{11} - t^{12} - t^{13} - t^{14} - t^{15})  $$
and $P(D_5)$ is as given in (I.1) for $D_5$ Lie algebra.

For $HA_2$, we obtain
$$ P(HA_2) \equiv { P(D_4) \over P_{11}(t) } \eqno(II.8) $$
where
$$ P_{11}(t) \equiv (1 - t) \ (1 + t)^3 \ (1 + t^2) \ (1 - t + t^2) \ (1 - t - t^3) . $$
As is emphasized above, we see that (II.8) has two other equivalent forms which can be expressed by
$$ P(HA_2) \equiv { P(A_3) \over P_5(t) } \eqno(II.9) $$
where
$$ P_5(t) \equiv (1 - t^2) \ (1-t-t^3)  $$
and

$$ P(HA_2) \equiv { P(A_4) \over P_9(t) } \eqno(II.10) $$
where
$$ P_9(t) \equiv (1 - t^2) \ (1-t-t^3) \ (1+t+t^2+t^3+t^4) . $$
One can see however that (II.7) does not allow us to get an equivalent
expression in the form
$$ P(HA_3) = { P(A_4) \over P(t) } $$
or
$$ P(HA_3) = { P(A_5) \over P^\prime(t) } $$
in the sense that P(t) or $P^\prime(t)$ can not be chosen as a polinomial of any finite 
order. This hence explains our maximality statement emphasized above. Let us also note 
that neither $P_{19}(t)$ in (II.7) nor $P_{11}(t)$ in (II.8) are completely factorizable 
in indeterminate t.

\vskip3mm
\noindent{\bf {REFERENCES}}
\vskip3mm

\item [1] V.Kac, Infinite Dimensional Lie Algebras
\item \ \ \ \ Cambridge University Press, 1982
\item [2] J.E.Humphreys, Introduction to Lie Algebras and Representation Theory 
\item \ \ \ \ Springer-Verlag, 1972
\item [3] J.E.Humphreys, Reflection Groups and Coxeter Groups, 
\item \ \ \ \ \ Cambridge University Press, 1990
\item [4] V.G.Kac, R.V.Moody and M.Wakimoto, On $E_{10}$
\item [5] H.R.Karadayi and M.Gungormez, On Weyl Character Formula and Signatures (work in preparation)

\end